\documentclass[fleqn,10pt]{wlscirep}

\newcommand{\grad}{\ensuremath{^\circ}}

\newcommand{\lsco}{La$_{2-x}$Sr$_{x}$CoO$_{4}$}
\newcommand{\prca}{Pr$_{2-x}$Ca$_{x}$CoO$_{4}$}

\newcommand{\lscoDD}{La$_{5/3}$Sr$_{1/3}$CoO$_{4}$}

\newcommand{\lscoV}{La$_{1.5}$Sr$_{0.5}$CoO$_{4}$}
\newcommand{\lscoVI}{La$_{1.4}$Sr$_{0.6}$CoO$_{4}$}
\newcommand{\lscoVII}{La$_{1.3}$Sr$_{0.7}$CoO$_{4}$}
\newcommand{\lscoVIII}{La$_{1.2}$Sr$_{0.8}$CoO$_{4}$}

\newcommand{\highest}{$\sim$90\%\ }
\newcommand{\highestx}{$0.9$}

\title{Incommensurate spin correlations in highly oxidized cobaltates \lsco}

\author[1]{Z. W. Li }
\author[1]{Y. Drees }
\author[1]{C. Y. Kuo }
\author[1]{H. Guo }
\author[2]{A. Ricci }
\author[3,4]{D. Lamago }
\author[5,6]{O. Sobolev }
\author[2]{U. R\"{u}tt}
\author[2]{O. Gutowski }
\author[7]{T. W. Pi }
\author[8]{A. Piovano }
\author[8,9]{W. Schmidt }
\author[1]{K. Mogare }
\author[1]{Z. Hu }
\author[1]{L. H. Tjeng }
\author[1,*]{A. C. Komarek }
\affil[1]{Max-Planck-Institute for Chemical Physics of Solids, N\"{o}thnitzer Str. 40, 01187 Dresden, Germany}
\affil[2]{Deutsches Elektronen-Synchrotron DESY, Notkestr. 85, 22603 Hamburg, Germany}
\affil[3]{Forschungszentrum Karlsruhe, Institut f\"ur
Festk\"orperphysik, P.O.B. 3640, D-76021 Karlsruhe, Germany}
\affil[4]{Laboratoire L\'{e}on Brillouin, CEA/CNRS,F-91191
Gif-sur Yvette Cedex, France}
\affil[5]{Forschungsneutronenquelle Heinz Maier-Leibnitz (FRM-II), TU M\"{u}nchen, Lichtenbergstr. 1,  D-85747 Garching, Germany}
\affil[6]{Georg-August-Universit\"{a}t G\"{o}ttingen, Institut f\"{u}r Physikalische Chemie, Tammannstrasse 6, D-37077 G\"{o}ttingen, Germany}
\affil[7]{National Synchrotron Radiation Research Center (NSRRC), Hsinchu 30077, Taiwan}
\affil[8]{Institut Laue-Langewin (ILL), 71 avenue des Martyrs, F-38042 Grenoble Cedex 9, France}
\affil[9]{J\"{u}lich Centre for Neutron Science JCNS, Forschungszentrum J\"{u}lich GmbH, Outstation at ILL, 71 avenue des Martyrs, F-38042 Grenoble Cedex 9, France}
\affil[*]{Alexander.Komarek@cpfs.mpg.de}

\begin{abstract}
We observe quasi-static incommensurate magnetic peaks in neutron scattering experiments on layered cobalt oxides \lsco\ with high Co oxidation states that have been reported to be paramagnetic.
This enables us to measure the magnetic excitations in this highly hole-doped incommensurate regime and compare our results with those found in the low-doped incommensurate regime that exhibit hourglass magnetic spectra. The hourglass shape of magnetic excitations completely disappears given a high Sr doping. Moreover, broad low-energy excitations are found, which are not centered at the incommensurate magnetic peak positions but around the quarter-integer values that are typically exhibited by excitations in the checkerboard charge ordered phase.
Our findings suggest that the strong inter-site exchange interactions in the undoped islands are critical for the emergence of hourglass spectra in the incommensurate magnetic phases of \lsco.
\end{abstract}
\begin{document}

\flushbottom
\maketitle
%
%
\thispagestyle{empty}

\section*{Introduction}

The recent observation of an hourglass magnetic spectrum in single-layer perovskite cobaltates \cite{hourglass} has attracted substantial attention because it resembles the hourglass-shaped excitation spectra that are universally observed in high-temperature superconducting (HTSC) cuprates \cite{hourglassLSCO,hourglassLSCOb,hourglassLSCOc,hourglassLSCOd,hourglassLSCOe,hourglassLSCOf,hourglassLSCOg,hourglassLSCOh,hourglassLSCOi,pseudo}.
This implies that a Fermi surface is not necessary in order for an hourglass spectrum to occur.
Because isostructural nickelates demonstrate the most robust diagonal charge stripe order at 1/3 hole doping, the incommensurate magnetic peaks in \lscoDD\ could be attributed to the presence of (disordered) charge stripe phases \cite{hourglass, musr,gaw,vojta}. In other words, this suggests a connection between hourglass spectra and charge stripe phases.
However, Y.~Drees \emph{et al.} have recently determined that essentially no (disordered) charge stripe phases exist in \lsco\ (1/3~$\leq$~$x$~$<$~1/2) \cite{drees, dreesB}. Therefore, the presence of incommensurate magnetic peaks was attributed to the frustration effects that arise in the disordered phases with strong residual character of the extremely stable checkerboard charge ordering (CBCO) correlations and to the presence of undoped Co$^{2+}$ islands in a nano phase separation scenario \cite{drees,dreesB,zli,guo}. It should be noted that the CBCO temperature in \lscoV\ is significantly high, i.e., 750--825~K  \cite{zz,half,cwik,reviewCo}, and that the CBCO correlations are consequently substantially stable in these cobaltates. For completeness, we would like to remark that in an earlier study, the small deviations from commensurate magnetism in La$_{1.5}$Sr$_{0.5}$CoO$_4$ have been attributed to the presence of stacking faults \cite{halfdopedinc}.

\par In order to determine which interactions are essential for the emergence of hourglass magnetic spectra, a study of cobaltates above half-doping (instead of below \cite{drees,dreesB,zli,guo}), is necessary. Although an incommensurate magnetic regime was found in the related cobaltate material \prca\ \cite{prca}, only a paramagnetic regime has been reported in \lsco\ for $x$~$>$~$0.6$ \cite{itoh,reviewCo,muSRB}. Here we observed a continuation of incommensurate magnetism across the half-doping and incommensurate magnetic peaks in the entire hole-doped regime of \lsco\ above the half-doping level (1/2~$<$~$x$~$\leq$~\highestx).
Thus, we were able to measure the previously unknown spin excitations of highly hole-doped cobaltates and to observe the differences between the low and high-doping regimes, which consequently allowed us to identify the strong exchange interactions in the undoped regimes as the fundamental elements of the presence of the high-energy portion of the hourglass spectra.

\section*{Results}

\par
\begin{figure}[!t]
\begin{center}
\includegraphics*[width=0.8\columnwidth]{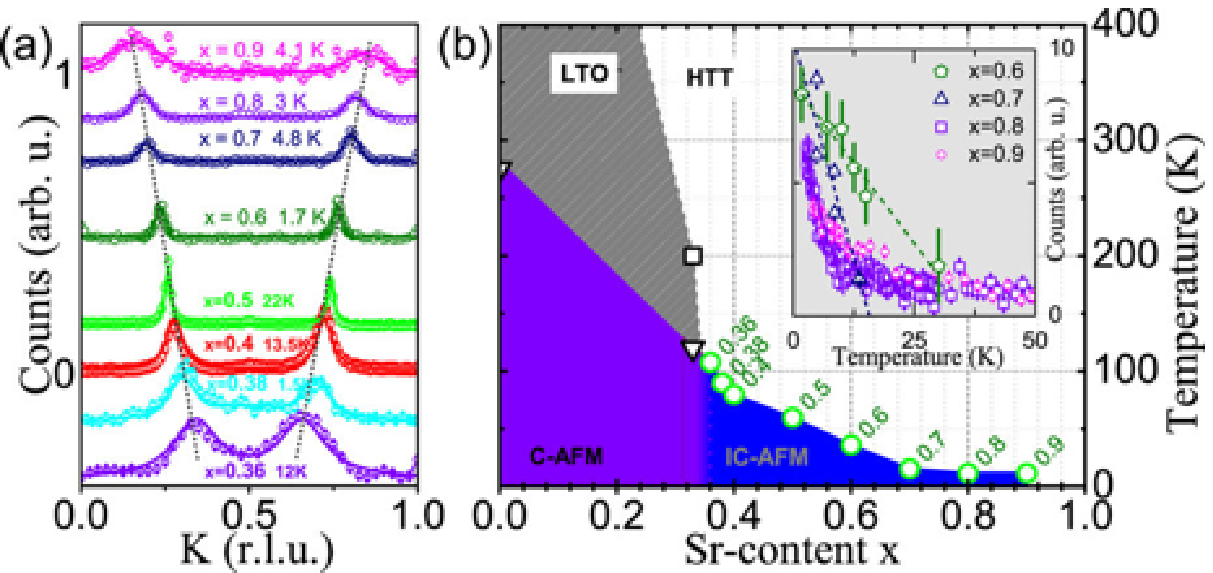}
\end{center}
\caption{(a) Incommensurate magnetic peaks observed in \lsco\ within neutron scattering experiments at the 1T, 2T, IN3 and PUMA spectrometers. Diagonal ($1$$-$$K$ $K$ $0$) scans across the magnetic satellites were performed. The dotted lines follow the magnetic peak positions and act as guide to the eyes. The solid lines are fitted to the data. Sr-concentrations $x$ and the measurement temperature are indicated. (b) Our \lsco\ phase diagram.  Transition temperatures are the onset-temperatures based on the temperature-dependent neutron measurements of the elastic neutron scattering intensities shown in the inset. The data points for $x$~$\leq$~0.5 were taken from \cite{drees}.}
\label{fig1}
\end{figure}

 Fig.~\ref{fig1}~(a) shows the neutron scattering intensities observed in the diagonal scans across the incommensurate magnetic peak positions, which were measured at low temperatures for \lsco\ with $x$~=~0.7, 0.8, and \highestx\, together with half-doped and lower-doped reference samples in order to create a comparison.
The entire highly hole-doped regime of \lsco\ clearly exhibits quasi-static incommensurate magnetic peaks in the elastic scans with neutrons.
The highest observed incommensurability amounts to $2\cdot\varepsilon_{mag}$~=~0.676(9) for our studied sample with \highest\ hole concentration. The incommensurability $\varepsilon$ is defined as two times the magnetic incommensurability $\varepsilon_{mag}$ such that magnetic reflections can be found at ($\pm\varepsilon_{mag}$ $\pm\varepsilon_{mag}$ 0) away from the planar antiferromagnetic wavevector which has half-integer $H$ and $K$ values. The charge incommensurability $\varepsilon_{charge}$ is defined as the incommensurability of its propagation vector ($\varepsilon_{charge}$ $\varepsilon_{charge}$ $L$). This value exceeds the nominal value of 0.5 for the half-doped checkerboard charge ordered sample which has not been observed that clearly before and shows that there is no saturation of the incommensurability around the value for CBCO.
Hence, the incommensurate magnetic regime below the half-doping level (which exhibits incommensurabilities $2\cdot\varepsilon_{mag}$ less than 1/2) naturally extends to the highly hole-doped regime (in which incommensurabilities $2\cdot\varepsilon_{mag}$ exceed 1/2). Our magnetic phase diagram, which is shown in Fig.~\ref{fig1}~(b), differs from Refs.~\cite{itoh,cwik,reviewCo,muSRB} where a paramagnetic regime above x~=~0.6 has been reported.

\par
\begin{figure}[!t]
\begin{center}
\includegraphics*[width=0.57\columnwidth]{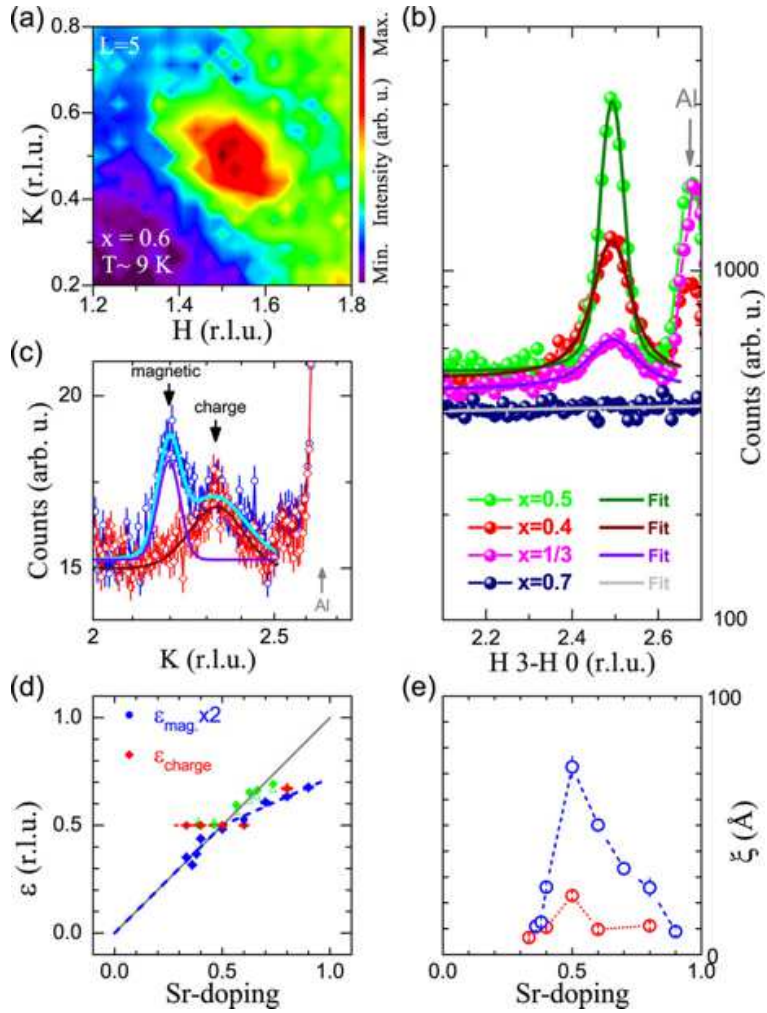}
\end{center}
\caption{(a) Synchrotron radiation single crystal X-ray diffraction measurements reveal half-integer peaks in \lscoVI\ at $\sim$9~K which are indicative for CBCO. (b) Neutron diffraction measurement of the charge correlations in \lscoVII. For direct comparison practically the same experimental conditions as for the study of the half- and underdoped cobaltates with $x$~$=$~0.5, 0.4 and 1/3 published in Ref.~\cite{guo} were reproduced. (c) Neutron scattering intensities measured in a diagonal ($3$$-$$K$~$K$~0) scan for the sample with $x$~$=$~0.8 at $\sim$3~K (blue) and 100~K (red). As can be seen, the charge correlations are incommensurate at such high hole-doping. The magnetic and charge ordering peak positions at 3~K were determined by fitting two Gaussians (cyan dotted line), and they equal 2.196(3) and 2.320(15), respectively. The charge ordering peak position at 100~K (solid magenta line) is 2.329(5). The corresponding incommensurabilities for charge and magnetic peaks are $2\cdot\varepsilon_{mag}$~=~0.608(6), $\varepsilon_{charge}$(3~K)~=~0.680(15), and $\varepsilon_{charge}$(100~K)~=~0.671(5). (d) Sr-doping dependence of the charge and magnetic incommensurabilities, $2\cdot\varepsilon_{mag}$ and $\varepsilon_{charge}$, respectively. Green data points are the corresponding values for \prca\ taken from Ref.~\cite{prca}; open/closed symbols: $2\cdot\varepsilon_{mag}$/$\varepsilon_{charge}$. (e) The correlation lengths $\xi_{||\mathbf{\widetilde{q}}}$ (\emph{blue}: magnetic; \emph{red}: charge) for different values of $x$. Here, the $\xi_{\perp\mathbf{\widetilde{q}}}$ are greater than $\xi_{||\mathbf{\widetilde{q}}}$ (e.g. 59(19)\%\ greater for $x$~=~0.7). It should be noted that $\xi_{\perp\mathbf{\widetilde{q}}}$ and $\xi_{||\mathbf{\widetilde{q}}}$ are defined as the correlation lengths perpendicular and parallel, respectively, to the direction of  $\mathbf{\widetilde{q}}$ with incommensurate magnetic peak positions $\mathbf{\widetilde{q}}$~=~$\mathbf{Q_{inc}}$~$-$~(1/2~1/2~0) and $\mathbf{Q_{inc}}$, respectively.
The correlation lengths were obtained from the inverse of the (Lorentzian) peak widths. For better visibility the charge correlation lengths have been multiplied by a factor $\times$3. The dashed lines act as guide to the eyes. The data for $x$~$\leq$~$0.5$ were taken from Refs.~\cite{guo,drees,dreesB}.  }
\label{fig5}
\end{figure}

 We studied the charge correlations in  \lscoVI, \lscoVII\  and \lscoVIII\ by means of single crystal X-ray and neutron diffraction.
The 60\%\ Sr-doped cobaltate is still within the CBCO regime since half-integer superstructure reflections can be observed in reciprocal space, see Fig.~\ref{fig5}~(a).
At 70\%\ Sr-doping the charge correlations are already that weak and diffuse that we were not able to detect them, see Fig.~\ref{fig5}~(b).
At even higher hole-doping the situation starts to change. In Fig.~\ref{fig5}~(c) elastic neutron scattering experiments on \lscoVIII\ are shown. A weak structural signal, which is indicated by an arrow in Fig~\ref{fig5}~(c), and that remains at higher temperatures (up to room temperature) is indicative for incommensurate charge correlations whereas the observed values of charge and magnetic incommensurabilities $2\cdot\varepsilon_{mag} \neq \varepsilon_{charge}$ seem to be incompatible with a simple charge stripe picture, see Fig.~\ref{fig5}~(d).
In \prca, qualitatively similar observations have been interpreted in terms of a glassy charge ordered state, which consists of multiple fragments \cite{prca}. Locally, these fragments share the common feature that any two Co$^{2+}$ ions are separated by a Co$^{3+}$ ion which is also the characteristic feature of checkerboard charge ordered islands within a disordered checkerboard charge ordered state as introduced in Ref.~\cite{dreesB}.
Also, in \prca\, these glassy charge correlations persist up to room temperature \cite{prca}.

\par
\begin{figure}[!t]
\begin{center}
\includegraphics*[width=0.57\columnwidth]{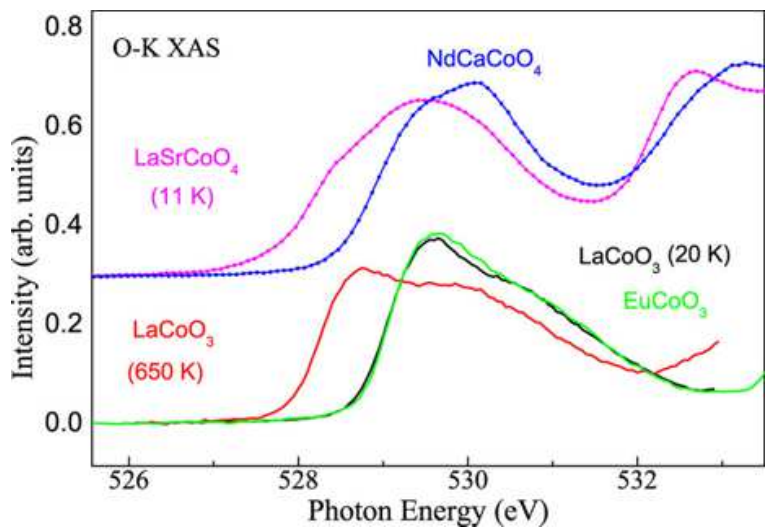}
\end{center}
\caption{O-K X-ray absorption spectra of LaSrCoO$_4$ at 11~K (magenta), NdCaCoO$_4$ at 300~K (blue), LaCoO$_3$ at 650~K, and 20~K (red and black) and EuCoO$_3$ at 300~K (green). }
\label{fig6}
\end{figure}

 The incommensurability in highly Sr-doped \lsco\ is smaller than that for similar Ca-doping in the \prca\ system \cite{prca}. We attribute this to the presence of a non-zero population of the Co$^{3+}$ high-spin (HS) state (S=2) in \lsco\ ($x$~$>$~1/2). Moreover, the smaller size of the Ca ions in \prca\ stabilizes the nonmagnetic Co$^{3+}$ low-spin (LS) state (S=0). In contrast to the small non-magnetic Co$^{3+}$ LS ions, the S=2 Co$^{3+}$ HS ions may contribute to the magnetism. Hence, when considering magnetism (magnetic exchange interactions) and structure (ionic sizes), the larger Co$^{3+}$ HS ions could effectively perform a role similar to the Co$^{2+}$ HS ions. In order to confirm this, we performed XAS studies, which are sensitive to the Co$^{3+}$ spin state. Fig.~\ref{fig6} displays the O-K edge spectrum of LaSrCoO$_4$ at 11~K (magenta), together with that of NdCaCoO$_4$ (at 300~K, blue), LaCoO$_3$ at 650~K (red), LaCoO$_3$ at 20~K (black), and EuCoO$_3$ (at 300~K, green). The NdCaCoO$_4$ and EuCoO$_3$ systems serve as references for materials that have Co$^{3+}$ in the pure LS state. Clearly, LaSrCoO$_4$ has its spectral features at energies that are lower than those of NdCaCoO$_4$, which is similar to those of LaCoO$_3$ at 650~K when compared with LaCoO$_3$ at 20~K and EuCoO$_3$. From earlier studies, we know that LaCoO$_3$ undergoes a gradual transition from LS to HS when going from 20~K to 650~K \cite{xasA,xasC}. Therefore, the LaSrCoO$_4$ spectrum at 11~K provides direct evidence that LaSrCoO$_4$ is not in a pure Co$^{3+}$ LS state. Instead, it contains non-negligible amounts of Co$^{3+}$ ions in the HS state. We estimate that LaSrCoO$_4$ has $\sim$30\% HS ions at low temperatures while La$_{1.5}$Sr$_{0.5}$CoO$_4$ has essentially no HS at low temperatures \cite{zz,chang}.
Finally, we also provide a XAS study of \lscoVIII\ which confirms that there is a significant population of the Co3+ HS state also at 80\% Sr-doping, see Supplementary Fig.~S1.

\par
\begin{figure}[!t]
\begin{center}
\includegraphics*[width=0.57\columnwidth]{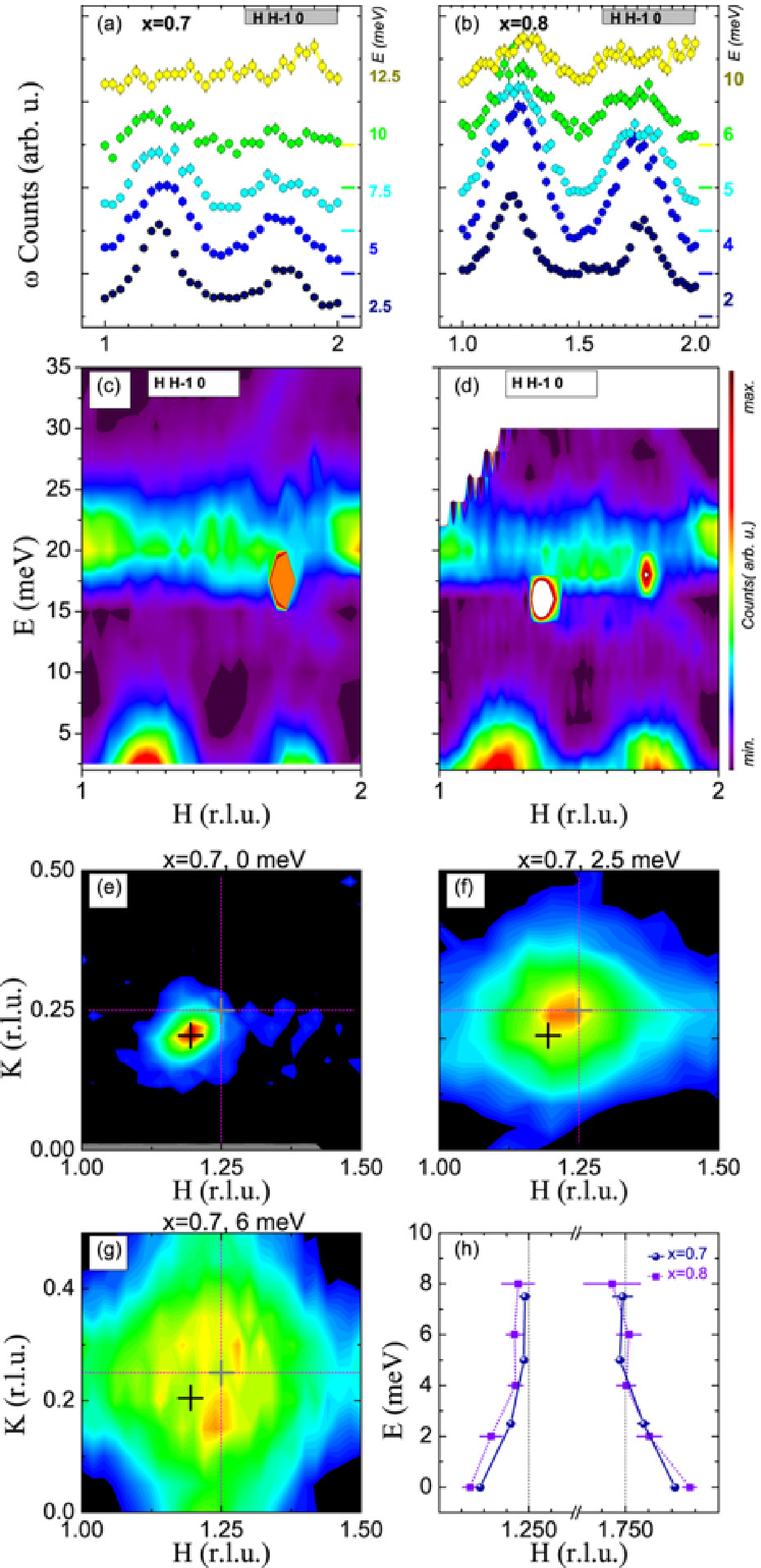}
\end{center}
\caption{Magnetic excitation spectra of \lscoVII, \lscoVIII. (a,b) Diagonal constant-E scans for different energies. (c,d) Neutron scattering intensities as a function of energy and momentum transfer in the color contour plots. (e--g) Magnetic correlations of \lscoVII\ in the $H$$K$$0$ plane of the reciprocal space for energy transfers of 0, 2.5, and 6~meV. The black and grey crosses indicate the positions of the incommensurate magnetic satellites in \lscoVII\ and in an ideal checkerboard charge-ordered \lscoV\ sample, respectively. (h) The center of the magnetic excitations in \lscoVII\ and \lscoVIII\ as a function of energy. }
\label{fig3}
\end{figure}

 The appearance of an incommensurate magnetic regime in highly hole-doped \lsco\ enables us to study its excitation spectra and compare the results with those gained from the lower Sr-doped regime \cite{hourglass,drees,dreesB}. Therefore, we measured the spin excitations of the 70\% and 80\% Sr-doped cobaltate at the 2T and PUMA spectrometers, respectively, and presented our results in Fig.~\ref{fig3}. As can be seen, the low-energy excitations are extremely broad when compared with those in the half-doped reference sample \cite{drees,half}. Furthermore, these excitations quickly vanish as energy increases.
(The near dispersionless $\sim$20~meV intensity band is most likely an optical phonon mode.)

\par
\begin{figure}[!t]
\begin{center}
\includegraphics*[width=0.7\columnwidth]{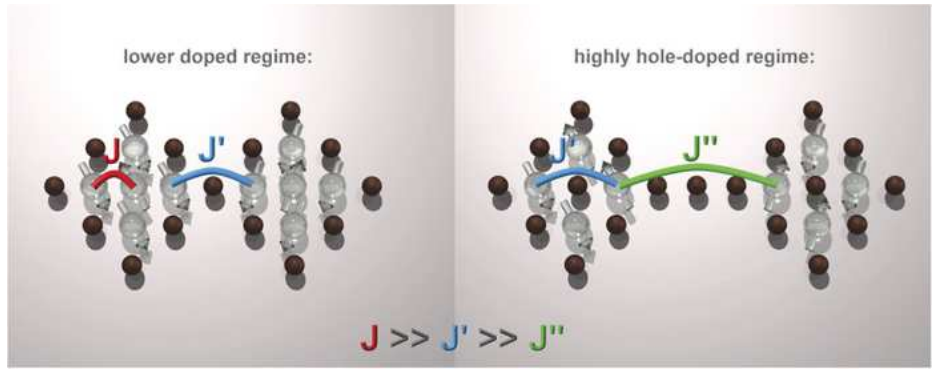}
\end{center}
\caption{Simplified schematic presentation of the exchange interactions and emergence of frustration in \lsco\ induced by electron (left side) or hole doping (right side). The light and dark spheres represent magnetic Co$^{2+,HS}$ and non-magnetic Co$^{3+,LS}$-ions, respectively. Exchange interactions $J$ (between nearest neighbors), $J'$ (across a hole), and $J''$ (across multiple holes) are shown. Only $J'$ appears in an ideal checkerboard charge ordered matrix. If additional electrons or holes are doped into this matrix (left and right images, respectively), $J'$ is replaced with $J$ or $J''$, respectively. This replacement induces frustration.}
\label{fig4}
\end{figure}

 In Fig.~\ref{fig3}~(e--g), an elastic intensity map and constant-energy maps are displayed for the low-energy magnetic excitations. When compared with the elastic signal (denoted "+" in Fig.~\ref{fig3}), it is apparent that these excitations are centered at positions in reciprocal space, displaced from the incommensurate magnetic peak positions of \lscoVII\ and much closer to the typical commensurate positions of the checkerboard charge ordered sample. The same observations can be seen for \lscoVIII\ in Fig.~\ref{fig3}~(h). This may indicate that the excitations for $x$~$\geq$~0.7 arise from the regions that have a short-ranged CBCO character, possibly occurring as isolated clusters given the high hole-doping level. This is similar to our recently reported nano phase separation scenario for the lower-doped regime of \lsco\ \cite{dreesB} with the difference that Co$^{2+}$ undoped islands are absent and the CBCO islands assume the role of these islands with strong exchange interactions, which can be seen in Fig.~\ref{fig4}. Moreover, these regions or fragments with CBCO character (i.e., the alternating Co$^{2+}$ -- Co$^{3+}$ paths along the $a$ or $b$ direction) may order in a manner that ensures the overall propagation vector is not half integer.
The substantially broad low-energy excitations, which are centered close to the quarter-integer positions in reciprocal space, also support the notion of isolated regions with short-ranged or local CBCO character. Similar to our observations, a glassy charge ordered state was already proposed for \prca\ in \cite{prca} (see Fig.~2(c) therein).

\section*{Discussion}
We found that the highly hole-doped regime of \lsco, up to the highest studied hole-doping level, is not entirely paramagnetic but exhibits quasi-static incommensurate magnetic correlations. We, therefore, revised the \lsco\ phase diagram.
During the review we noticed that a $\mu$SR study was carried out on the magnetic phase diagram of \lsco\ \cite{muSRB}. The phase diagram there is different due to the fact that there is oxygen non-stoichiometry in the samples used in the $\mu$SR study \cite{muSRB}. Nevertheless, for the high Sr regime the $\mu$SR study \cite{muSRB} observes that the spin fluctuations slow down upon cooling, not inconsistent with our neutron observations of quasi-static spin correlations regarding the different time scales probed in neutron (few ps) and $\mu$SR ($\sim\mu$s) measurements \cite{muSRB} (or also in NMR measurements \cite{itoh}).

\begin{figure}[!t]
\begin{center}
\includegraphics*[width=0.57\columnwidth]{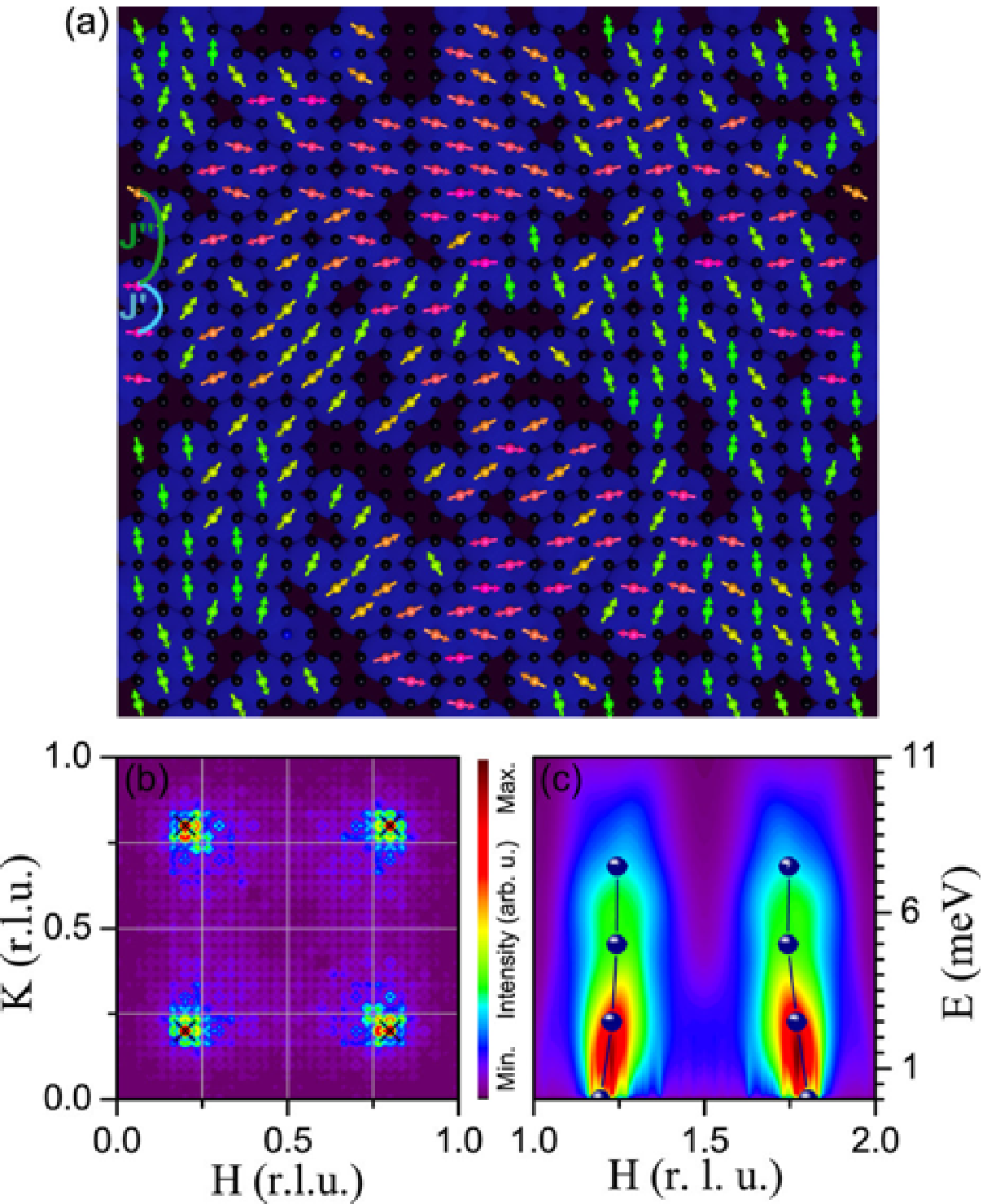}
\end{center}
\caption{(a) Spin structure calculated for $x$~$\sim$~0.7. The black spheres represent Co$^{3+}$ LS ions, and the colored spheres with the arrows indicate Co$^{2+}$ HS ions with spins. (b)  Corresponding magnetic neutron scattering intensities in the $HK0$ plane: \emph{crosses} signify experimental peak positions for $x$~=~0.7. (c) Magnetic excitation spectrum together with the experimentally observed values for this Sr doping (circles). }
\label{fig7}
\end{figure}

Whereas the lower-doped regime of \lsco\ exhibits magnetic excitation spectra with an hourglass shape, the hourglass shape of magnetic excitations completely disappears in the highly hole-doped regime. This might indicate that the hourglass spectrum in \lsco\ requires the presence of undoped islands with Co$^{2+}$-Co$^{2+}$ exchange interactions and clarifies the very recent findings in Refs.~\cite{drees,dreesB}.
In order to address this issue, we carried out spin-wave simulations for 70\% Sr-doped material using the \emph{McPhase} program code \cite{mcp}. The Monte-Carlo and spin-wave simulations were constructed as in Ref.~\cite{dreesB}. Here, we used a hole probability of 70\% which is at the border between experimentally still clearly observeable CBCO correlations ($x$~$=$~0.6) and incommensurate charge correlations ($x$~$=$~0.8). The steric repulsion of two large adjacent Co$^{2+}$-ions was the dominating contribution creating the simulated charge configurations.
This most simple model is able to create disordered CBCO correlations, see Supplementary~Fig.~S2.
In addition to the parameters in Ref.~\cite{dreesB}, we also included exchange interactions $J(300)$~=~-0.15~meV and $J(400)$~=~-0.1~meV between Co$^{2+}$ ions, with a distance of three and four unit cells, respectively, i.e.,  $J(100)$~$\equiv$~$J$~=~-5.8~meV, $J(200)$~$\equiv$~$J'$~=~-0.85~meV, $J(300)$~=~-0.15~meV, and $J(400)$~$\equiv$~$J''$~=~-0.1~meV. For the easy plane anisotropy of each Co$^{2+}$ ion, we assumed an in-plane magnetic saturation moment of $m_x=m_y=2\mu_B\langle-|\hat S_x|+\rangle=2i\mu_B\langle-|\hat S_y|+\rangle=3~\mu_B$ and an out-of-plane saturation moment of $m_z=2\cdot\mu_B\langle+|\hat S_z|+\rangle = 2\cdot 1.3 \mu_B = 2.6 \mu_B$.
In Fig~\ref{fig7}~(a) the calculated spin structure is shown.
Our results indicate that non-collinear magnetic structures appear due to frustration. Moreover, the corresponding magnetic peak positions strongly resemble the experimental observations, as can be seen by comparing Fig.~\ref{fig7}~(b) with Fig.~\ref{fig3}~(e).

Hence, our simulations support also a nano phase separation scenario for the highly hole-doped regime of \lsco\ at least up to $x$~$\sim$~0.7.
Unlike for the lower-doped regime, in which both nano phase separated regions are magnetic, at higher hole-doping the nano phase separated regions primarily consist of non-magnetic Co$^{3+}$ LS clusters and regions with local CBCO character (blue and black areas in Fig~\ref{fig7}~(a)).
Furthermore, the calculated magnetic excitation spectrum resembles the experimental observations, which is demonstrated by Fig.~\ref{fig7}~(c). A more detailed view into the calculated excitations reveals that for the low-energy part (e.g. at 1.305~meV) Co-ions coupled with all kind of exchange interactions contribute to the strongest spin excitations whereas at higher energies (e.g. at 8.37~meV) basically only some clusters with Co-ions that are coupled with the strongest existing exchange interactions $J'$ contribute to the excitations, see supplementary video files, i.e. these excitations are basically hosted within Co$^{2+}$-Co$^{3+}$-Co$^{2+}$ charge fragments which we would like to call regions with CBCO-like character.
This corroborates our nano phase separation scenario further and indicates why the excitations become centered around quarter-integer peak positions with increasing energies (which is typical for excitations of Co ions coupled with $J'$ in the ideal CBCO ordered cobaltate).

Although the magnetic peaks appear to be incommensurate at high hole-doping in \lsco, the hourglass spectrum is absent if the nanoscopic undoped islands with strong exchange interactions $J$ are not present in our model.
Consequently, the highly hole-doped regime of \lsco\ reveals excitations, from which we can conclude that the strong inter-site exchange interactions within the nano phase separated undoped regions are essential for the emergence of hourglass spectra.

\section*{Methods}

\lsco\ single crystals have been grown under high oxygen pressures ($>$9~bar) using the floating zone technique, following the same synthesis method applied to lower Sr-doped cobaltates \cite{growth}. In order to ensure the absence of (tiny) oxygen deficiencies, the sample with $x$~=~0.8 was also post-annealed for seven~days at 400~\grad C under an oxygen pressure of 5000~bar. Nevertheless, no apparent increase in the incommensurabilities was observed when compared with other similarly grown single crystals.
Elastic and inelastic neutron scattering experiments have been performed using 1T, 3T.1, and 2T triple-axis spectrometers at the Laboratoire L\'{e}on Brillouin (LLB) in Saclay, France, as well as the PUMA spectrometer at the FRM-II in Garching, Germany and the IN3 spectrometer at the ILL in Grenoble, France. Higher-order contributions were suppressed by the usage of two pyrolythic graphite (PG) filters in the experiments.
Synchrotron radiation single crystal X-ray diffraction measurements were performed at beamline P07 using 100~keV hard X-rays.
The soft X-ray absorption spectroscopy (XAS) at the O-K edge of LaSrCoO$_4$, NdCaCoO$_4$, LaCoO$_3$, and EuCoO$_3$ was measured using a photon energy resolution of 0.25~eV at the NSRRC in Taiwan.


\section*{Acknowledgements (not compulsory)}

We thank Y.~Sidis for his help and discussions,
 the team of H.~Borrmann for the powder X-ray diffraction measurements,
 the team of U.~Burkhardt for the EDX and WDX measurements,
 the team of G.~Auffermann for the ICP measurements,
 the team of M.~Schmidt for the TG measurements,
and A.~Todorova for her assistance.

\section*{Author contributions statement}

Conceiving experiments: A. C. K.; conducting and analyzing experiments: Z. W. L., Y. D., H. G., C. Y. K., A. R., D. L., O. S., U. R., O. G., T. W. P., A. P., W. S., K. M., Z. H. and A. C. K.; spin wave simulations: A. C. K.; interpretation and manuscript writing: A.~C.~K. and L. H. T.

\section*{Additional information}

\textbf{Competing financial interests} none.

\end{document}